# DASEE: A Synthetic Database of Domestic Acoustic Scenes and Events in Dementia Patients' Environment

Abigail Copiaco[1], Christian Ritz[2], Stefano Fasciani[3], and Nidhal Abdulaziz[4]
University of Wollongong, Australia[1, 2]
University of Wollongong in Dubai, UAE[1, 4]
University of Oslo, Norway[3]
E-mail/s: {abigailc, critz}@uow.edu.au, stefano.fasciani@imv.uio.no, {abigailcopiaco, nidhalabdulaziz}@uowdubai.ac.ae
Dubai Knowledge Park, P.O. Box 20183, Dubai, UAE, Fax: +971 4 2781801

*Abstract* – Access to informative databases is a crucial part of notable research developments. In the field of domestic audio classification there have been significant advances in recent years. Although several audio databases exist, these can be limited in terms of the amount of information they provide, such as the exact location of the sound sources, and the associated noise levels. In this work, we detail our approach on generating an unbiased synthetic domestic audio database, consisting of sound scenes and events, emulated in both quiet and noisy environments. Data is carefully curated such that it reflects issues commonly faced in a dementia patient's environment, and recreate scenarios that could occur in real-world settings. Similarly, the room impulse response generated is based on a typical one-bedroom apartment at Hebrew SeniorLife Facility. As a result, we present an 11-class database containing excerpts of clean and noisy signals at 5-seconds duration each, uniformly sampled at 16 kHz. Using our baseline model using Continues Wavelet Transform Scalograms and AlexNet, this yielded a weighted F1-score of 86.24%.

## I. INTRODUCTION

Dementia, a neurodegenerative ailment experienced by the elderly, is commonly associated with cognitive decline [1]. Due to its progressive nature, it affects how the patient perceives external stimuli, especially noise and light [2]. Hence, patients may experience distress, provided that they perceive external stimuli differently compared to those unaffected by dementia. Such distress may result in wandering, and changes in behaviour [3,4]. For these reasons, consistent monitoring is crucial for maintaining a safe environment for the dementia patient. Monitoring systems are commonly used as a form of assistive technology to help inform carers of the patients' assistance requirement. However, visual monitoring systems are often subject to concerns in privacy infringement [5,6]. Therefore, audio-based systems make a sufficient alternative, provided that microphones are less intimidating when compared to cameras.

The Sound Interfacing through the Swarm (SINS) database is a domestic acoustic scene database, which is composed of 9 different recording categories, sampled at 16 kHz [7]. Although this is sufficient to conduct an initial experiment, and to test the effectiveness of the proposed system in terms of sound classification, there are several limitations to this, especially when conducting an in-depth analysis of the system performance. Firstly, the data provided publicly are extracted solely from the first four receiver nodes of the 13-node setup [7]. Each node consists of a linear array four microphones with individual channel sampling. The four nodes are also located only along the living room and kitchen areas, which extensively limits the scope of the area. Further, it also poses a challenge on covering sound categories that are deemed dangerous to the patients. Similarly, the exact locations of the sound sources are not provided in the database.

Synthesizing a new database allows including data that address these issues and include additional disruptive sounds commonly faced in a dementia patient's environment. Further, it also allows to recreate scenarios that could occur in real-world settings, including noisy environments, and various source-to-receiver distances. Furthermore, this will also provide the exact locations of the sound sources, which will be useful for sound location estimation purposes.

The structure of the paper is organized such that Section II provides details about the data sample curation, experimental setup, and room impulse response generation. Section III then discusses the process of synthesizing and refining the database, eliminating the potential of biasing and overfitting. Section IV describes the approach used to integrate noise into the clean signals, creating a more realistic database. Finally, the concluding section gives suggestions to improve and extend the scope for future work.

## II. SYNTHETIC DATA GENERATION

The developed database promotes domestic acoustic events and scenes likely to happen in a realistic dementia patient care facility. The following sections provide detailed information regarding this.

### A. Data Curation

Monitoring disruptive noises for dementia patients can be challenging, even for healthcare professionals, as sound levels acceptable to staff may be distressing for dementia patients. This is due to the fact that dementia may worsen the effects of sensory changes, as the progressive nature of this ailment may alter how the patient perceives external stimuli, such as acoustic noise pollution [8]. A summary of possible negative impacts caused by disruptive noise levels to dementia patients is provided below:

- As hearing is linked to balance, aural disruption could lead to greater risks of falls, either through loss of balance [9], or through an increase in disorientation as a result of people trying to orientate themselves in an overstimulating environment [8].

- It has been proven that dementia patients respond on a more sensory level, rather than intellectually. For example, they note the body language or tone of voice, rather than what people actually say [10]. Since people with dementia have a reduced ability to understand their sensory environment, when combined with age-related deterioration in hearing, it can be overwhelming.

- Other research suggests that wandering behaviour in dementia patients may be their way to try to remove themselves from an overstimulating situation [11].

The following are examples of disruptive noises that patients normally experience [8]:

- **Sudden noises:** such as toilet flushes, alarms, glass shattering.
- **Unnecessary noises:** such as television that is not being watched, people talking, and loud music. Eliminating unnecessary noise can reduce the risks of aggression in noisy environments.
- **Sounds in open spaces:** some sounds appear louder in open spaces, for example, noises from a kitchen and dining area, the wheels of a tea trolley or the sound of conversations or laughing.
- **Inappropriate noise timing:** acoustic noise pollution at night can result in disturbed sleep which in turn can lead to problems during the day, such as lack of concentration, and difficulty communicating and performing during the day.

Considering these factors, Table 1 summarizes the sources of the dry sample excerpts from which this database was generated using, along with the types of sounds extracted from them.

Table 1. Dry Sample Sources and Licensing Summary.

| Database | Categories Used | License |
|---|---|---|
| DESED Synthetic Soundscapes [12] | Alarm, Blender, Frying, Shaver, Water | MIT Open Source Initiative |
| Kaggle: Audio Cats & Dogs [13] | Cat, Dog | CC BY-SA 3.0 |
| Open SLR: 64 and 70 [14] | Speech (Marathi Multi-speaker, English) | CC BY-SA 4.0 |
| FSDKaggle2019 [15] | Scream, Slam, Shatter | CC BY-4.0 |
| FSD50K [16] | Dishes | CC BY-4.0 |
| SINS Database [7] | Absence / Silence | CC BY-NC 4.0 |
| UrbanSound8k [17] | Background Noises | CC BY-NC 3.0 |

*B. Experimental Setup*

The generation of the synthetic database is based on a 999 square-foot one-bedroom apartment in Hebrew Senior Life Facility [18], illustrated in Figure 1, where the dimensions are mentioned in meters.

We assume a 3-m height for the ceiling of the apartment. Multi-channel recordings are aimed for; hence, node receivers are placed on every four corners of each of the six rooms concerned, at 0.2 m below the ceiling. Each node receiver is a microphone array composed of four linearly arranged omnidirectional microphones with 5 cm inter-microphone spacing, as per the geometry detailed in Figure 2.

The room dimensions, source and receiver locations, wall reflectance, and other relevant information were used in order to compute the impulse responses for each room. These are then convoluted with the dry sounds, specifying their location, in order to create the synthetic data.

Details regarding the process of sound synthesis are provided in the succeeding sub-section of this paper.

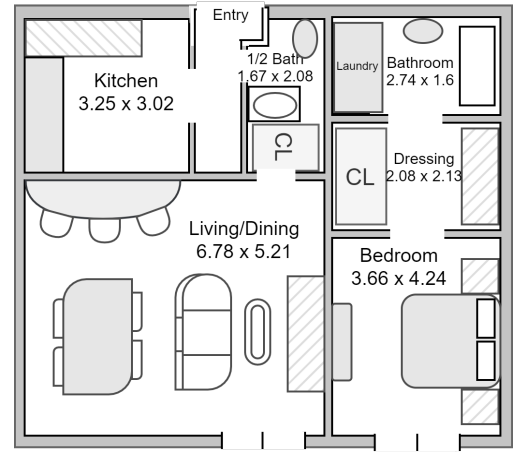

Fig 1 One-bedroom apartment in Hebrew SeniorLife Facility [18].

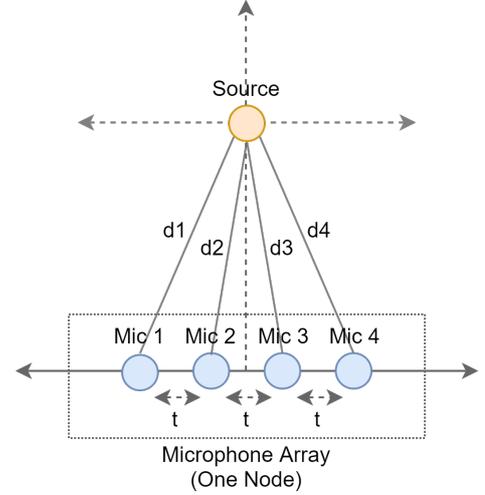

Fig 2 Microphone Array Geometry for a single node: Four linearly spaced microphones.

Table 1. Room Dimensions and Source/Receiver Locations for Multi-channel Recordings in the DASEE database

| Room | Dimension | Sounds | Source Locations | Node | Receiver Locations |
|---|---|---|---|---|---|
| **Bedroom** | [3.6576, 4.2418, 3] | Alarm | [3.30, 1.50, 0.5] | 1 | [3.45, 0.05, 2.8], [3.5, 0.05, 2.8], [3.55, 0.05, 2.8], [3.6, 0.05, 2.8] |
| | | Speech | [2, 2.5, 1.6], [1, 1, 1.8] | | |
| | | TV | [0.25, 2.121, 1.7] | | |
| | | Shatter | [3, 1.5, 0.25] | 2 | [3.45, 4.2, 2.8], [3.5, 4.2, 2.8], [3.55, 4.2, 2.8], [3.6, 4.2, 2.8] |
| | | Scream | [2, 2.10, 1] | | |
| | | Slam | [0.2, 4, 2], [1.5, 0.25, 2] | 3 | [0.05, 0.05, 2.8], [0.1, 0.05, 2.8], [0.15, 0.05, 2.8], [0.2, 0.05, 2.8] |
| | | Silence | [2, 2, 1] | 4 | [0.05, 4.2, 2.8], [0.1, 4.2, 2.8], [0.15, 4.2, 2.8], [0.2, 4.2, 2.8] |
| **Living or Dining Room** | [6.7818, 5.207, 3] | Cat | [3, 2, 0.4], [5, 4, 0.4], [4, 2.5, 0.4] | 1 | [6.58, 0.05, 2.8], [6.63, 0.05, 2.8], [6.68, 0.05, 2.8], [6.73, 0.05, 2.8] |
| | | Dog | [3, 2, 0.4], [5, 4, 0.4], [4, 2.5, 0.4] | | |
| | | TV | [6.59, 2.10, 1.5] | 2 | [6.58, 5.16, 2.8], [6.63, 5.16, 2.8], [6.68, 5.16, 2.8], [6.73, 5.16, 2.8] |
| | | Speech | [3.5, 2.5, 1.6], [1.2, 2.5, 1.8] | | |
| | | Scream | [3.35, 2, 1] | | |
| | | Silence | [3, 2.5, 1] | 3 | [0.05, 0.05, 2.8], [0.1, 0.05, 2.8], [0.15, 0.05, 2.8], [0.2, 0.05, 2.8] |
| | | Slam | [6, 5, 2], [6.4, 0.25, 2] | 4 | [0.05, 5.16, 2.8], [0.1, 5.16, 2.8], [0.15, 5.16, 2.8], [0.2, 5.16, 2.8] |
| | | Shatter | [1.2, 2.5, 1], [1.2, 5, 1] | | |
| **Kitchen** | [3.2512, 3.0226, 3] | Blender | [0.25, 2.80, 1.2] | 1 | [3.05, 0.05, 2.8], [3.1, 0.05, 2.8], [3.15, 0.05, 2.8], [3.2, 0.05, 2.8] |
| | | Dishes | [1.1, 0.05, 1] , [1, 0.5, 1] | 2 | [3.05, 2.93, 2.8], [3.1, 2.93, 2.8], [3.15, 2.93, 2.8], [3.2, 2.93, 2.8] |
| | | Frying | [0.25, 1.85, 1] | 3 | [0.05, 0.05, 2.8], [0.1, 0.05, 2.8], [0.15, 0.05, 2.8], [0.2, 0.05, 2.8] |
| | | Scream | [0.4, 2.5, 1.6] | | |
| | | Shatter | [1.75, 1.5, 1] | 4 | [0.05, 2.93, 2.8], [0.1, 2.93, 2.8], [0.15, 2.93, 2.8], [0.2, 2.93, 2.8] |
| **Bath** | [2.7432, 1.6002, 3] | Tooth brush | [0.6, 0.4, 1] | 1 | [2.54, 0.05, 2.8], [2.59, 0.05, 2.8], [2.64, 0.05, 2.8], [2.69, 0.05, 2.8] |
| | | Scream | [2, 1.4, 1] | 2 | [2.54, 1.55, 2.8], [2.59, 1.55, 2.8], [2.64, 1.55, 2.8], [2.69, 1.55, 2.8] |
| | | Shatter | [0.6, 0.35, 1] | 3 | [0.05, 0.05, 2.8], [0.1, 0.05, 2.8], [0.15, 0.05, 2.8], [0.2, 0.05, 2.8] |
| | | Water | [0.6, 0.4, 1], [2.55, 1.55, 0.65] | 4 | [0.05, 1.55, 2.8], [0.1, 1.55, 2.8], [0.15, 1.55, 2.8], [0.2, 1.55, 2.8] |
| **Half-bath** | [1.6764, 2.0828, 3] | Tooth brush | [0.05, 0.05, 1] | 1 | [1.47, 0.05, 2.8], [1.52, 0.05, 2.8], [1.57, 0.05, 2.8], [1.62, 0.05, 2.8] |
| | | Scream | [0.85, 1.04, 1.6] | 2 | [1.47, 2.03, 2.8], [1.52, 2.03, 2.8], [1.57, 2.03, 2.8], [1.62, 2.03, 2.8] |
| | | Shatter | [0.05, 0.15, 1] | 3 | [0.05, 0.05, 2.8], [0.1, 0.05, 2.8], [0.15, 0.05, 2.8], [0.2, 0.05, 2.8] |
| | | Water | [0.05, 0.05, 1] | 4 | [0.05, 2.03, 2.8], [0.1, 2.03, 2.8], [0.15, 2.03, 2.8], [0.2, 2.03, 2.8] |
| **Dressing Room** | [2.0828, 2.1336, 3] | Speech | [1.04, 1.05, 1.8] | 1 | [1.88, 0.05, 2.8], [1.93, 0.05, 2.8], [1.98, 0.05, 2.8], [2.03, 0.05, 2.8] |
| | | | | 2 | [1.88, 2.08, 2.8], [1.93, 2.08, 2.8], [1.98, 2.08, 2.8], [2.03, 2.08, 2.8] |
| | | Silence | [1, 1, 1] | 3 | [0.05, 0.05, 2.8], [0.1, 0.05, 2.8], [0.15, 0.05, 2.8], [0.2, 0.05, 2.8] |
| | | | | 4 | [0.05, 2.08, 2.8], [0.1, 2.08, 2.8], [0.15, 2.08, 2.8], [0.2, 2.08, 2.8] |

## C. Room Impulse Response Generation

Data curation is achieved through convoluting the dry sound excerpts with the relevant room impulse response generated. The impulse response were synthesized at a sampling rate of 16 kHz, using linear interpolation method with azimuth elevation source orientation [19]. Sounds with fixed position source, such as flowing water through a sink, and alarm clock placed on the bed; are convoluted in a single location. On the other hand, sounds with a movable source, such as speech and animal sounds, are convoluted with the room impulse response at up to three different locations. All relevant information regarding the room dimensions, as well as source and receiver locations, are provided in Table 2.

The wall reflection coefficients utilized in the convolution process also vary on the room locations, depending on the percentage of obstruction by furniture, and whether it is a regular wall, floor, or ceiling. Table 3 enlists the average room reflectance depending on the percentage of walls that are obstructed, using common wall reflectance coefficients [20]. Similarly, according to the European Standard EN 12464, ceilings have a typical wall reflectance coefficient of 0.7-0.9, walls have 0.5-0.8, and floors have 0.2-0.4 [21].

Table 3. Average room reflectance for varying wall reflectance and obstruction percentages [20].

| Walls Obstruct | Wall Reflectance | | | | | | |
|---|---|---|---|---|---|---|---|
| | 0.2 | 0.3 | 0.4 | 0.5 | 0.6 | 0.7 | 0.8 |
| 20% | 0.475 | 0.505 | 0.535 | 0.565 | 0.596 | 0.626 | 0.656 |
| 30% | 0.488 | 0.515 | 0.541 | 0.569 | 0.594 | 0.620 | 0.647 |
| 40% | 0.502 | 0.524 | 0.547 | 0.570 | 0.592 | 0.615 | 0.638 |
| 50% | 0.515 | 0.534 | 0.553 | 0.572 | 0.591 | 0.610 | 0.628 |
| 60% | 0.529 | 0.544 | 0.559 | 0.574 | 0.589 | 0.604 | 0.619 |
| 70% | 0.542 | 0.553 | 0.565 | 0.576 | 0.587 | 0.599 | 0.610 |
| 80% | 0.555 | 0.563 | 0.571 | 0.578 | 0.586 | 0.593 | 0.601 |

According to these guidelines, taking into consideration the wall type and obstruction percentages, the wall reflectance coefficients utilized in the setup for the generation of the room impulse response are seen in Table 4. The same wall reflectance coefficient was used for the four sides of the walls, and different coefficients were used for the ceiling, and the floor, as per the European Standard EN [20].

Table 4. Wall reflectance coefficients used to synthesize the DASEE database

| Room | Reflectance | Obstruction | Reflectance Utilized |
|---|---|---|---|
| Bedroom | Walls – 0.5<br>Ceiling – 0.7<br>Floor – 0.2 | Walls – 30, 50, 70, 30<br>Ceiling – 0<br>Floor – 30 | Walls – 0.568, 0.572, 0.576, 0.568<br>Ceiling – 0.7<br>Floor - 0.488 |
| Living or Dining | Walls – 0.5<br>Ceiling – 0.7<br>Floor – 0.2 | Walls – 30, 50, 50, 20<br>Ceiling – 0<br>Floor – 30 | Walls – 0.568, 0.572, 0.572, 0.568<br>Ceiling – 0.7<br>Floor – 0.488 |
| Kitchen | Walls – 0.6<br>Ceiling – 0.8<br>Floor – 0.3 | Walls – 30, 30, 30, 30<br>Ceiling – 0<br>Floor – 20 | Walls – 0.594, 0.594, 0.594, 0.594<br>Ceiling – 0.8<br>Floor – 0.515 |
| Bath | Walls – 0.7<br>Ceiling – 0.8<br>Floor – 0.4 | Walls – 20, 30, 0, 0<br>Ceiling – 0<br>Floor – 30 | Walls – 0.626, 0.62, 0.7, 0.7<br>Ceiling – 0.8<br>Floor – 0.541 |
| Half-bath | Walls – 0.7<br>Ceiling – 0.8<br>Floor – 0.4 | Walls – 20, 20, 0, 0<br>Ceiling – 0<br>Floor – 30 | Walls – 0.626, 0.626, 0.7, 0.7<br>Ceiling – 0.8<br>Floor – 0.541 |
| Dressing Room | Walls – 0.5<br>Ceiling – 0.7<br>Floor – 0.2 | Walls – 80, 80, 20, 20<br>Ceiling – 0<br>Floor – 30 | Walls – 0.578, 0.578, 0.565, 0.565<br>Ceiling – 0.7<br>Floor - 0.488 |

## III. DATA SYNTHESIS AND REFINEMENT

Excerpts from the source databases vary in duration, number of channels, and sampling frequency. For example, the DESED database and Freesound data use a sampling frequency of 44.1 kHz, can be single or dual channels, and may vary in duration [12]. On the other hand, excerpts from the SINS database are sampled at 16 kHz, are recorded from four channels, and 10-s in duration [7]. Although the DESED database also included mixed data, only single source excerpts were used in the process of synthesizing this database. For our DASEE database all signal channels are averaged prior to being resampled to 16 kHz. After this, they are subject to a six-step synthesis and refinement method, as summarized in Figure 3.

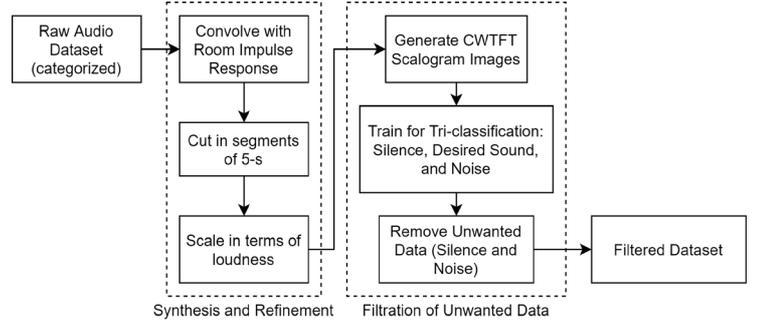

Fig 3 DASEE Database Synthesis and Refinement Process.

As illustrated in Figure 3, raw audio data is first convoluted with the relevant room impulse responses generated per channel, before being concatenated upon writing, creating a four-channel data. All excerpts are then cut into segments with 5-s duration each in order to promote uniformity. Finally, these are then scaled in terms of loudness by adding white noise at an ideal signal-to-noise ratio for a noise-free environment, which is 80 dB [22].

However, since longer durations are chopped into segments of 5-s, some of these segments are not guaranteed to contain the desired sound event. Therefore, a neural network-based filtration method is utilized in order to remove unwanted audio files. Excerpts of 1000 audio files that do not contain sound events and scenes are categorized as 'Silence', while 1000 audio files that contain desired sounds are labelled as 'Desired Sounds'. Lastly, another set containing 1000 files is categorized with the label 'Noise'. Through this, a three-level classifier was developed through the CWTFT scalograms and CNN method via AlexNet pre-trained network. Through our previous work, this combination was found to provide accurate results for domestic acoustic classification [23, 24]. This network is then used to classify the entire synthesized database. Only those that fell under the 'Desired Sounds' category are kept, and those that fall under the two are filtered out as misclassified data.

## IV. BACKGROUND NOISE INTEGRATION

In order to reflect a realistic environment, along with the clean set, recordings with background acoustic noise pollution are also included in the database. For the background noise, we used excerpts from the Noise Urban Sound 8K database [17]. This composes of 8732 labelled sound excerpts of urban sounds that are coming from 10 different classes, including: air conditioner, car horn, children playing, dog barking, drilling, engine idling, gun shot, jackhammer, siren, and street music [17]. Data from the noise sounds that are more relevant for a dementia patient's environment is selected, which includes: air conditioner, children playing, and street music. Aside from this, white noise is also added as background noise for some of the files.

To add the background acoustic noise matching the synthetically generated database, the noise files are also resampled from 44 kHz to 16 kHz and sliced to 5 seconds segments. Accordingly, they are also convoluted with the relevant room impulse responses before being added to the clean data. The air conditioning background noise was assumed to be placed near the walls, elevated slightly lower than the ceiling, while noises such as "children playing" and "street music" were placed near open windows. Background noise were added at different Signal-to-Noise (SNR) levels, including: 15 dB, 20 dB, and 25 dB, in order to reflect real life recordings. This work utilizes the Voicebox MATLAB toolbox in order to add the noise at specified SNR levels [25].

*A. Curating an Unbiased Database*

The data generated must be curated into verified training and testing sets, ensuring that several instances of the testing data do not exist within the training data. This is done in order to overcome the risks of overfitting the network, which occurs when the network performs exceedingly well solely for the particular database that it was trained for.

Although splitting the data randomly between the training and the testing set can help overcome this potential, it may still be risky due to the several instances of each sound that occur in the database, as per Figure 4. The database developed contains recordings from the four different nodes across each room. Further, there are 4 instances of these from the addition of the noises for the same sound at 3 different SNR levels. Hence, the curation of the data into training and testing sets follow the algorithm in the figure below.

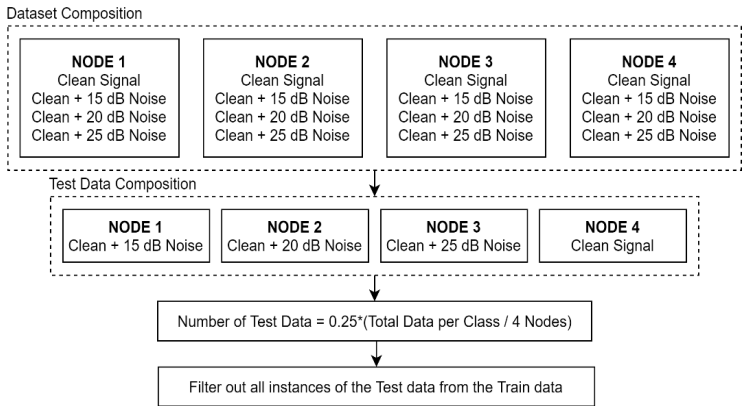

Fig 4 DASEE database Training and Testing Data Curation Process.

As illustrated in Figure 4, the training and testing sets were constructed in such a way that it avoids the chances of overfitting. In particular, each node is assigned with one of each specific noise levels, while the fourth node is assigned with the clean signal. This ensures that all four instances will have significant differences in them. Similarly, they will also have different noise levels, as it would in real life recordings, where a certain noise can be closer to a single node compared to the other nodes in the same room.

Following this curation process, the summary of the new database generated can be seen in Table 5. For this, all instances of any recording that exists in the test set has been removed from the training set. It is important to note that smaller categories such as: "Dishes" and "Frying", have been combined into one folder called "Kitchen Activities". This database filtration helps gets rid of potential biasing and overfitting, which will be helpful for the robustness of our training results. Furthermore, this also helps get a fairer comparison between different networks, provided that they all use the same training and testing set, instead of a randomized split.

Table 5 Summary of the Training and Testing Data.

| Category | Training Data | Testing Data |
|---|---|---|
| Absence / Silence | 11286 | 876 |
| Alarm | 2765 | 260 |
| Cat | 11724 | 1080 |
| Dog | 6673 | 792 |
| Kitchen_Activities | 12291 | 1062 |
| Scream | 4308 | 376 |
| Shatter | 2877 | 370 |
| Shaver_toothbrush | 11231 | 1077 |
| Slam | 1565 | 268 |
| Speech | 30113 | 2374 |
| Water | 6796 | 829 |
| **TOTAL** | **101629** | **9364** |

## V. RESULTS

In this section, we provide a per-level and overall results report on the dataset presented in this paper using a baseline technique. This approach uses the Continuous Wavelet Transform (CWT) Scalograms as features to the AlexNet pre-trained model, which provided us accurate results, as identified in our previous work [24]. The results for this is summarized in Table 6.

The performance of the approach was assessed through several performance evaluation metrics, inclusive of the Accuracy, Precision, Recall, and F1-scores. Further, the use of weighted, micro, and macro F1-score averaging was used to take into consideration the imbalance throughout the dataset developed. The slight inconsistency in the performance figures observed throughout the classes is due to the presence of both sound events and acoustic scenes in the dataset.

## VI. CONCLUSIONS

This paper details the process we used to generate a synthetic domestic acoustic scene and event database that contains sounds commonly experienced in a dementia patient's environment. The database covers 11 sound categories, namely: Absence, Alarm, Cat, Dog, Kitchen Activities, Scream, Shatter, Shaver / toothbrush, Slam (such as the door slamming), Speech, and Water. All audio files are uniformly sampled at 16 kHz, has a 5-second duration, and are four-channel in nature. Aside from sound classification purposes, this database may also be used for sound source location estimation. The database is released publicly in order to be utilized for future research.

Table 6 Per-level and Overall Results Summary

| Category | Train | Test | TP | FP | FN | Accuracy | Precision | Recall | F1-score |
|---|---|---|---|---|---|---|---|---|---|
| Absence | 11286 | 876 | 876 | 2 | 0 | 100.00% | 99.77% | 100.00% | 99.89% |
| Alarm | 2765 | 260 | 168 | 86 | 92 | 64.62% | 66.14% | 64.62% | 65.37% |
| Cat | 11724 | 1080 | 1054 | 193 | 26 | 97.59% | 84.52% | 97.59% | 90.59% |
| Dog | 6673 | 792 | 580 | 34 | 212 | 73.23% | 94.46% | 73.23% | 82.50% |
| Kitchen | 12291 | 1062 | 878 | 385 | 184 | 82.67% | 69.52% | 82.67% | 75.53% |
| Scream | 4308 | 376 | 317 | 88 | 59 | 84.31% | 78.27% | 84.31% | 81.18% |
| Shatter | 2877 | 370 | 289 | 58 | 81 | 78.11% | 83.29% | 78.11% | 80.61% |
| Shaver | 11231 | 1077 | 765 | 224 | 312 | 71.03% | 77.35% | 71.03% | 74.06% |
| Slam | 1565 | 268 | 178 | 54 | 90 | 66.42% | 76.72% | 66.42% | 71.20% |
| Speech | 30113 | 2374 | 2374 | 42 | 0 | 100.00% | 98.26% | 100.00% | 99.12% |
| Water | 6796 | 829 | 608 | 111 | 221 | 73.34% | 84.56% | 73.34% | 78.55% |
| **TOTAL** | **101629** | **9364** | **8087** | **1277** | **1277** | **86.36%** | **86.36%** | **86.36%** | **86.36%** |
| | | | | | Weighted | 86.36% | 86.72% | 86.36% | 86.24% |
| | | | | | Macro | 81.03% | 82.99% | 81.03% | 81.69% |

# REFERENCES

[1] I. Korolev, "Alzheimer's Disease: A Clinical and Basic Science Review," *Medical Student Research Journal*, vol. 4, 2014, pp. 24-33.

[2] Social Care Institute for Excellence, *Dementia-friendly environments: Noise levels*, May 2015, Accessed on: Aug 20 2020, [online]. Available: https://www.scie.org.uk/dementia/supporting-people-with-dementia/dementia-friendly-environments/

[3] J. van Hoof, H.S.M. Kort, M.S.H. Duijnstee, P.G.S. Rutten, and J.L.M. Hensen, 'The indoor environment and the integrated design of homes for older people with dementia", *Building & Environment*, 45(5), 2010, pp. 1244–1261

[4] J.D. Price, D.G. Hermans, and J. Grimley Evans, "Subjective barriers to prevent the wandering of people cognitively impaired people", Cochrane Database of Systematic reviews, 3: CD001932, 2007..

[5] J. Cocco, "Smart home technology for the elderly and the need for regulation," *Journal of Environmental and Public Health Law*, 6(1), 2011, pp. 85-108

[6] B. Bennett, et al., "Assistive Technologies for People with Dementia: Ethical Considerations," *Bulletin of the World Health Organization*, 2017, pp. 1-12

[7] G. Dekkers et al., "The SINS database for detection of daily activities in a home environment using an acoustic sensor network," *DCASE*, 2017.

[8] Social Care Institute for Excellence , "Dementia-friendly environments: Noise levels", published May 2015, Accessible at: https://www.scie.org.uk/dementia/supporting-people-with-dementia/dementia-friendly-environments/noise.asp#:~:text=Of%20all%20the%20senses%2C%20hearing,such%20as%20noise%20and%20light, Accessed Aug 20, 2020.

[9] Hayne, M.J, and Fleming, R. "Acoustic design guidelines for dementia care facilities", Proceedings on 43rd International Congress on Noise Control Engineering: Internoise 2014, Melbourne, Australia, pp. 1-10.

[10] Van Hoof, J., Kort, H.S.M., Duijnstee, M.S.H., Rutten, P.G.S. and Hensen, J.L.M. 'The indoor environment and the integrated design of homes for older people with dementia', Building and Environment, 2010, vol 45, no 5, pp 1244–1261.

[11] Price, J.D., Hermans, D.G. and Grimley Evans, J. Subjective barriers to prevent the wandering of people cognitively impaired people, Cochrane Database of Systematic reviews, 3: CD001932, 2007.

[12] Turpault, N.; Serizel, R.; Shah, A.P., and Salamon, J. Sound event detection in domestic environments with weakly labeled data and soundscape synthesis. In Proc. DCASE Workshop, Oct 2019

[13] Takahashi, N.; Gygli, M.; Pfister, B.; and Van Gool, L. Deep Convolutional Neural Networks and Data Augmentation for Acoustic Event Recognition, Proc. Interspeech, San Fransisco, 2016.

[14] He, F. et al. Open-source Multi-speaker Speech Corpora for Building Gujarati, Kannada, Malayalam, Marathi, Tamil and Telugu Speech Synthesis Systems, Proceedings of The 12th Language Resources and Evaluation Conference (LREC), Marseille, France, 2020.

[15] Fonseca, E.; Plakal, M.; Font, F.; Ellis, D.P.W.; Serra, X. Audio tagging with noisy labels and minimal supervision. Proceedings of the DCASE 2019 Workshop, NYC, US, 2019.

[16] Fonseca, E.; Favory, X.; Pons, J.; Font, F.; and Serra, X. "FSD50K: an Open Dataset of Human-Labeled Sound Events", arXiv:2010.00475, 2020.

[17] Salamon, J.; Jacoby, C.; and Bello, J.P., A Dataset and Taxonomy for Urban Sound Research, 22nd ACM International Conference on Multimedia, Orlando USA, Nov. 2014.

[18] Hebrew SeniorLife. [Online]. Accessible at: https://www.hebrewseniorlife.org/newbridge/types-residences/independent-living/independent-living-apartments.

[19] Hafezi, S. Moore, A.H., and Naylor, P.A. Room Impulse Response for Directional source generator (RIRDgen), 2015. Accessible at: http://www.commsp.ee.ic.ac.uk/~ssh12/RIRD.htm

[20] Simm, S.; and Coley, D. The relationship between wall reflectance and daylight factor in real rooms, Architectural Science Review, vol. 54, no. 4, pp. 329-334, 2011.

[21] European Committee for Standardization, "EN 12464-1. Lighting of work places - Part 1: Indoor work places.," 2011.

[22] Boudreau, M., "What is Signal to Noise Ratio (SNR)? What to look for & how to use it", How to record a podcast, 12$^{th}$ May 2017. Accessible at: https://www.thepodcasthost.com/recording-skills/signal-to-noise-ratio/

[23] Copiaco, A.; Ritz, C.; Fasciani, S.; and Abdulaziz, N. Scalogram Neural Network Activations with Machine Learning for Domestic Multi-channel Audio Classification, Proceedings of the IEEE ISSPIT Conference, 2019, Ajman, United Arab Emirates, pp. 1-6.

[24] Copiaco, A.; Ritz, C.; Abdulaziz, N.; and Fasciani, S. Identifying Optimal Features for Multi-channel Acoustic Scene Classification, Proceedings of the ICSPIS Conference, 2019, Dubai, United Arab Emirates, pp. 1-4.

[25] Brookes, M., 'v_addnoise', Voicebox MATLAB Toolbox, Accessible at: http://www.ee.ic.ac.uk/hp/staff/dmb/voicebox/mdoc/v_mfiles/v_addnoise.html